\documentclass[aps,amsmath,notitlepage,twocolumn,amssymb,prl,longbibliography]{revtex4-1}

\usepackage{graphicx}
\usepackage{dcolumn}
\usepackage{bm}
\usepackage[colorlinks=true,citecolor=red,urlcolor=blue]{hyperref}
\usepackage{wasysym}
\usepackage{stmaryrd}
\usepackage{verbatim}
\usepackage{subfigure}
\usepackage{amsmath}

\begin{document}
\title{Magnetic competition induced colossal magnetoresistance in $\mathit{n}$-type Hg$\mathbf{Cr_{2}}$$\mathbf{Se_{4}}$ under high pressures}

\author{J. P. Sun$^{1,2}$}
\author{Y. Y. Jiao$^{1,3}$}
\author{C. J. Yi$^{1,2}$}
\author{S. E. Dissanayake$^{4,5}$}
\author{M. Matsuda$^{4}$}
\author{Y. Uwatoko$^{6}$}
\author{Y. G. Shi$^{1,2,7}$}
\author{Y. Q. Li$^{1,2,7}$}
\author{Z. Fang$^{1,2,7}$}
\author{J.-G. Cheng$^{1,2,7}$}
\email[Email: ]{jgcheng@iphy.ac.cn}
\affiliation{$^{1}$Beijing National Laboratory for Condensed Matter Physics and Institute of Physics, Chinese Academy of Sciences, Beijing 100190, China\\
	$^{2}$School of Physical Sciences, University of Chinese Academy of Sciences, Beijing 100190, China\\
	$^{3}$Faculty of Science, Wuhan University of Science and Technology, Wuhan, Hubei 430065, China\\
	$^{4}$Neutron Scattering Division, Oak Ridge National Laboratory, Oak Ridge, TN 37831, USA\\
	$^{5}$Department of Physics, Duke University, Durham, North Carolina, 27708, USA\\
	$^{6}$Institute for Solid State Physics, University of Tokyo, 5-1-5 Kashiwanoha, Kashiwa, Chiba 277-8581, Japan\\
	$^{7}$Songshan Lake Materials Laboratory, Dongguan, Guangdong 523808, China}

\date{\today}

\begin{abstract}
The $\mathit{n}$-type Hg$\mathrm{Cr_{2}}$$\mathrm{Se_{4}}$ exhibits a sharp semiconductor-to-metal transition (SMT) in resistivity accompanying the ferromagnetic order at $\mathit{T_{\rm C}}$ = 106 K. Here, we investigate the effects of pressure and magnetic field on the concomitant SMT and ferromagnetic order by measuring resistivity, dc and ac magnetic susceptibility, as well as single-crystal neutron diffraction under various pressures up to 8 GPa and magnetic fields up to 8 T. Our results demonstrate that the ferromagnetic metallic ground state of $\mathit{n}$-type Hg$\mathrm{Cr_{2}}$$\mathrm{Se_{4}}$ is destabilized and gradually replaced by an antiferromagnetic, most likely a spiral magnetic, and insulating ground state upon the application of high pressure. On the other hand, the application of external magnetic fields can restore the ferromagnetic metallic state again at high pressures, resulting in a colossal magnetoresistance (CMR) as high as $\mathrm{\sim 3 \times 10^{11}}$ $\%$ under 5 T and 2 K at 4 GPa. The present study demonstrates that $\mathit{n}$-type Hg$\mathrm{Cr_{2}}$$\mathrm{Se_{4}}$ is located at a peculiar critical point where the balance of competion between ferromagnetic and antiferromagnetic interactions can be easily tipped by the external stimuli, providing a new platform for achieving CMR in a single-valent system.
\end{abstract}

\maketitle

The chromium chalcogenide spinels Cd$\mathrm{Cr_{2}}$$\mathrm{S_{4}}$, Cd$\mathrm{Cr_{2}}$$\mathrm{Se_{4}}$, and Hg$\mathrm{Cr_{2}}$$\mathrm{Se_{4}}$ are well-known ferromagnetic semiconductors that have been studied for several decades~\cite{JAP1966,JAP1967,PRL1965,PR1966}. Many anomalous physical properties such as colossal magnetoresistance (CMR)~\cite{PSS1997,PSS2008}, anomalous Hall effect~\cite{PSS1997}, and the red shift of the optical absorption edge~\cite{JPSJ1973} have been observed in these compounds due to the intimated correlation between charge and spin degrees of freedom.

Recently, Hg$\mathrm{Cr_{2}}$$\mathrm{Se_{4}}$  has received renewed interest because the first-principles calculations have predicated a novel Chern semimetal state and quantized Anomalous Hall effect in this compound~\cite{PRL2011}. In the earlier reports, Hg$\mathrm{Cr_{2}}$$\mathrm{Se_{4}}$  was found to exhibit semiconducting behavior with a ferromagnetic (FM) transition taking place at $\mathit{T_{\rm C}}$ = 106 ${\sim}$ 120 K~\cite{PRL1965,PR1966,JAP1978}, which can be tuned by either hole or electron doping~\cite{JPSJ1971}. Its saturated magnetic moment 5.64 ${\mu_{\mathrm B}}$${/}$f.u. is close to the expected high-spin $\mathrm{Cr^{3^{+}}}$ ion~\cite{PRL1965,PR1966}. Following the theoretical prediction, some of us have synthesized high quality $\mathit{n}$-type Hg$\mathrm{Cr_{2}}$$\mathrm{Se_{4}}$ single crystals, for which the FM order at $\mathit{T_{\rm C}}$ = 106 K is accompanied with a very sharp semiconductor to metal transition (SMT)~\cite{JLTP2013,PRL2015}. The resistivity was found to drop eight orders of magnitude in the FM state below $\mathit{T_{\rm C}}$. Under an external magnetic field, the FM transition moves to higher temperatures, leading to a CMR near $\mathit{T_{\rm C}}$, e.g. $\mathrm{\sim 7 \times 10^{6}}$ $\%$ at 8 T and 110 K~\cite{PRL2015,PRB2016Lin}. Andreev reflection spectroscopy measurements provide direct evidence that $\mathit{n}$-type Hg$\mathrm{Cr_{2}}$$\mathrm{Se_{4}}$ is a half-metal below $\mathit{T_{\rm C}}$ with spin polarizations as high as 97$\%$~\cite{PRL2015}. Since the interesting properties of Hg$\mathrm{Cr_{2}}$$\mathrm{Se_{4}}$ occurs around or below $\mathit{T_{\rm C}}$ = 106 K, it is desirable to enhance $\mathit{T_{\rm C}}$ to higher temperatures from the viewpoint of practical applications.

High pressure, as an effective and clean knob, can be employed to precisely tune the crystal structure, electronic and magnetic properties of materials. A comprehensive high-pressure study on Hg$\mathrm{Cr_{2}}$$\mathrm{Se_{4}}$ single crystal, however, is still lacking. An earlier high-pressure study on Hg$\mathrm{Cr_{2}}$$\mathrm{Se_{4}}$ showed that its room-temperature resistivity decreases by one order of magnitude in the pressure range 0.5 - 1.5 GPa, followed by another weak anomaly at around 6 GPa~\cite{JLCM1984}. The dramatic drop of resistivity was explained as a pressure-induced semiconductor to metal transition~\cite{JLCM1984}. However, the metallic state around room temperature was not verified by the temperature dependence of resistivity under pressure. A recent theoretical calculation based on density-functional theory seems to support a pressure-induced semiconductor to metal transition, but the predicted critical pressure of 11.4 GPa is an order of magnitude higher~\cite{JPCM2012}. Efthimiopoulos $\mathit{et}$ $\mathit{al.}$ performed a high-pressure structural study and found two successive structural transformations in Hg$\mathrm{Cr_{2}}$$\mathrm{Se_{4}}$; i.e., the cubic $\mathit{Fd}$-3$\mathit{m}$ phase first transforms to a tetragonal $\mathit{I}$$4_{1}$${/}$$\mathit{amd}$ phase at about 15 GPa and then to a structurally disordered or partial amorphous phase above 21 GPa~\cite{APL2014}. It was speculated that the $\mathit{Fd}$-3$\mathit{m}$ to $\mathit{I}$$4_{1}$${/}$$\mathit{amd}$ transformation is accompanied with an insulator to metal transition because the Raman signal disappears after the structural transition~\cite{APL2014}. Therefore, all the existing experimental and theoretical studies indicated the possible occurrence of pressure-induced semiconductor/insulator to metal transition in Hg$\mathrm{Cr_{2}}$$\mathrm{Se_{4}}$, which has not been verified so far to the best of our knowledge.

Here we performed a comprehensive high-pressure study on $\mathit{n}$-type Hg$\mathrm{Cr_{2}}$$\mathrm{Se_{4}}$ single crystals by measuring its magnetotransport, dc and ac magnetic susceptibility, as well as single-crystal neutron diffraction under various pressures up to 8 GPa. Details about crystal growth and experimental methods are given in Supplementary materials. Surprisingly, we found the FM metallic ground state of $\mathit{n}$-type Hg$\mathrm{Cr_{2}}$$\mathrm{Se_{4}}$ is destabilized and gradually replaced by an antiferromagnetic (AF), most likely a spiral magnetic order, and insulating ground state upon the application of high pressure, which is counterintuitive and in strikingly contrast to all previous studies mentioned above~\cite{JLCM1984,JPCM2012,APL2014}. On the other hand, the application of external magnetic fields can restore the FM metallic state again at high pressures, resulting in a CMR as high as ${\sim 3 \times 10^{11}}$ $\%$ under 5 T and 2 K at 4 GPa. Our present study demonstrates that Hg$\mathrm{Cr_{2}}$$\mathrm{Se_{4}}$ situates at a critical point where the competition between FM and AF exchange interactions can be easily tuned by pressure and magnetic field. Due to the strong coupling between the spin and charge degrees of freedom, $\mathit{n}$-type Hg$\mathrm{Cr_{2}}$$\mathrm{Se_{4}}$ thus represents a unique example that the magnetic phase competition can induce a significant CMR in a single-valent system.

\begin{figure}[h]
	\centering
	\includegraphics[width=0.45\textwidth]{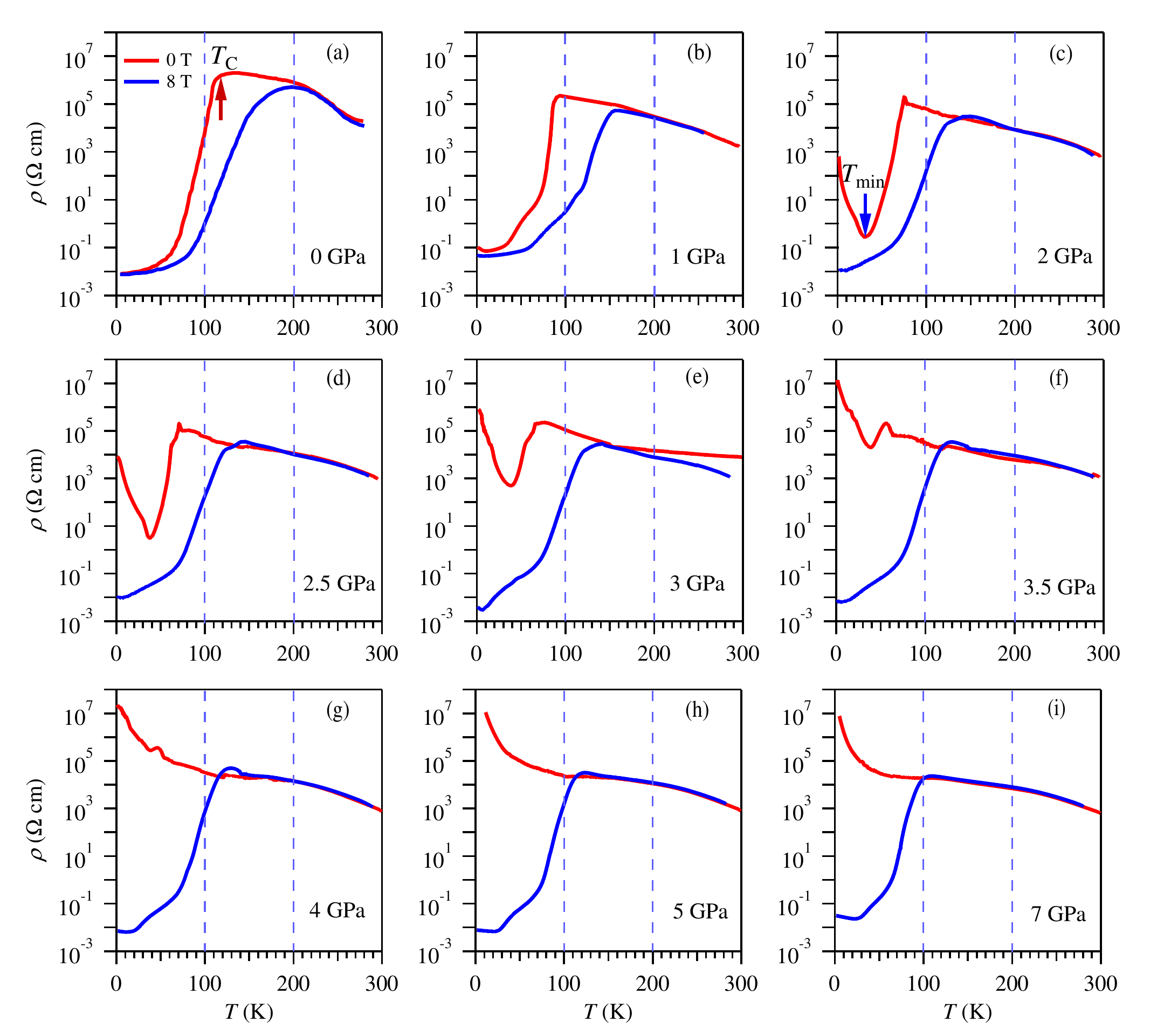}
	\caption{(color online) (a) - (i) Temperature dependence of resistivity $\mathit{\rho(\mathit{T})}$ of Hg$\mathrm{Cr_{2}}$$\mathrm{Se_{4}}$ at 0 T and 8 T under various pressures up to 7 GPa.}
	\label{Fig. 1}
\end{figure}

Fig.1 displays the temperature dependence of resistivity $\mathit{\rho(\mathit{T})}$ for $\mathit{n}$-type Hg$\mathrm{Cr_{2}}$$\mathrm{Se_{4}}$ at 0 T and 8 T under various pressures up to 7 GPa. At ambient pressure, its $\mathit{\rho}$$(\mathit{T})$ at 0 T displays a sharp SMT manifested by about eight orders drop in resistivity below the FM order at $\mathit{T_{\rm C}}$ = 106 K, Fig. 1(a). Here, we define $\mathit{T_{\rm C}}$ as the interception between two straight lines below and above. The SMT in $\mathit{\rho}$$(\mathit{T})$ at 8 T moves to higher temperature of $\sim$ 200 K, resulting in a CMR ${\sim 10^{6}}$ $\%$ around $\mathit{T_{\rm C}}$, in agreement with previous studies~\cite{PRL2015,PRB2016Lin}. When we increase pressure to 1 GPa, the sharp drop in resistivity is still very clear, but the FM ordering has been lowered to $\mathit{T_{\rm C}}$ $\sim$ 90 K, Fig. 1(b). Meanwhile, a weak upturn appears in $\mathit{\rho}$$(\mathit{T})$ below $\mathit{T_{\rm min}}$ $\sim$ 10 K. Similarly, the SMT at 1 GPa also shifts to higher temperature $\sim$ 150 K upon the application of external magnetic field of 8 T. Upon further increasing the pressure to 2 GPa, the concomitant FM order and the SMT were suppressed to $\mathit{T_{\rm C}}$ $\sim$ 70 K. Surprisingly, the low-temperature upturn in resistivity is strongly enhanced, i.e. the resistivity increases by three orders of magnitude below $\mathit{T_{\rm min}}$, which has also been enhanced to $\sim$ 30 K at 2 GPa. The opposite pressure dependencies of $\mathit{T_{\rm C}}$ and $\mathit{T_{\rm min}}$ result in the reduction of the metallic-like region in between $\mathit{T_{\rm C}}$ and $\mathit{T_{\rm min}}$.  Again, the application of 8 T magnetic field restores the metallic state, Fig. 1(c). With further increasing pressure to $\mathit{P}$ $\geq$ 2.5 GPa, the temperature $\mathit{T_{\rm min}}$ does not change too much, but the metallic-like region at $\mathit{T_{\rm min}}$ $<$ $\mathit{T}$ $<$ $\mathit{T_{\rm C}}$ shrinks quickly and almost vanishes at $\mathit{P}$ $>$ 3.5 GPa, when the semiconducting behavior retains to the lowest temperature. Such an observation of pressure-induced metal to insulator transition in $\mathit{n}$-type Hg$\mathrm{Cr_{2}}$$\mathrm{Se_{4}}$ is quite unexpected and is in strikingly contrast with all previous experimental reports and theoretical predictions mentioned above~\cite{JLCM1984,JPCM2012,APL2014}. Interestingly, we found that the application of magnetic field can change the situation dramatically. As shown in Fig. 1(d-i), the temperature dependence of resistivity at 8 T exhibits sharp SMT as observed at lower pressures. Our results thus indicate that high pressure destabilizes the FM metallic ground state, whereas magnetic field can restore it.

Before we proceed to investigate the underlying mechanism for such an intriguing phenomenon, we first evaluate the detailed effect of magnetic field on resistivity or MR under each pressure. Fig. 2 shows the representative $\mathit{\rho(\mathit{T})}$ and MR data under different magnetic fields at pressures of 2, 3, 4, and 7 GPa. Here we define MR ratio as 100 $\%$ $\times$ ($\mathit{\rho_\mathrm{xx,0}}$-$\mathit{\rho_\mathrm{xx}(\mathit{H})}$)$/$$\mathit{\rho_\mathrm{xx}(\mathit{H})}$. For $\mathit{P}$ $=$ 2 GPa, the resistivity upturn below $\mathit{T}_\mathrm{min}$ $\sim$ 30 K can be easily suppressed by a moderate magnetic field $<$ 0.5 T, above which the metallic state persists down to the lowest temperature. Here, we can clearly see that $\mathit{T}_\mathrm{min}$ is progressively suppressed by external magnetic field, and the MR increases from ${10^{2}}$ $\%$ to ${10^{6}}$ $\%$ upon cooling below $\mathit{T}_\mathrm{min}$. With increasing pressure to 3 GPa, the resistivity value at $\mathit{T}_\mathrm{min}$ is three orders of magnitude higher than that at 2 GPa, illustrating that pressure further stabilizes the insulating state. Accordingly, the external magnetic field that is needed to restore the metallic state also increases quickly with pressure, and a metallic state cannot be fully recovered up  to 8 T at 7 GPa. As seen in Fig. 2 (f, g, h), a colossal MR as high as $\sim$ $10^{11}$ $\%$ can be achieved under 5 T and 2 K at 4 GPa or under 8 T and 2 K at 7 GPa. Such a colossal MR $\sim$ $10^{11}$ $\%$ is close to the largest CMR values in the perovskite manganites~\cite{PRL1993}. These results further demonstrate that the external pressure and magnetic field play opposite roles in controlling the electronic ground state of this system.  

\begin{figure}[h]
	\centering
	\includegraphics[width=0.5\textwidth]{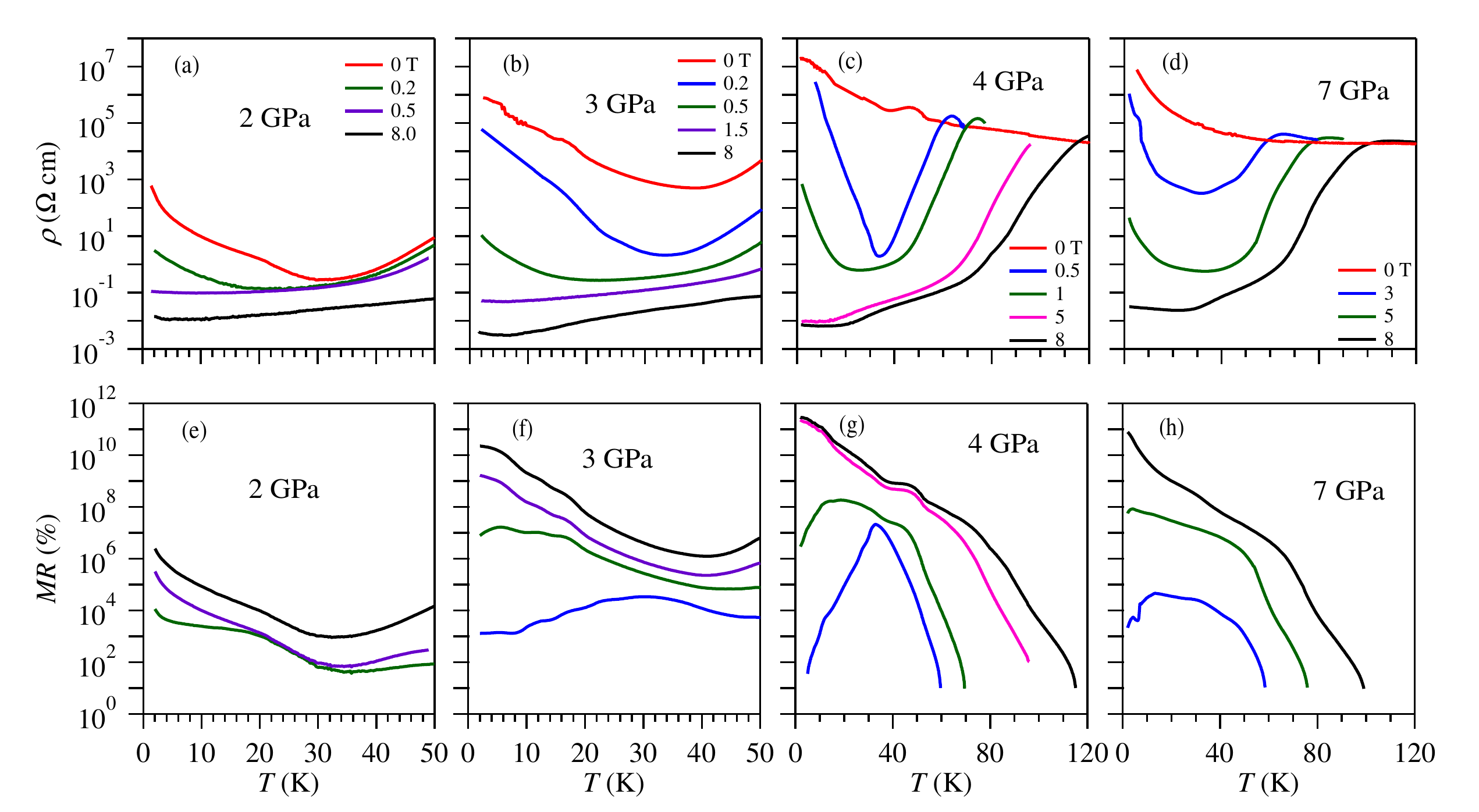}
	\caption{(color online) Temperature dependence of resistivity $\mathit{\rho(\mathit{T})}$ and MR ratio at different magnetic fields under various represented pressures: (a) 2 GPa, (b) 3 GPa, (c) 4 GPa, and (d) 7 GPa. In Fig. 2 (a) - (d) and (e) - (h), the resistivity $\mathit{\rho(\mathit{T})}$ and MR have the same coordinate range as in Fig. 2 (a) and (e).}
	\label{Fig. 2}
\end{figure}

Because the electrical transport properties of half-metallic Hg$\mathrm{Cr_{2}}$$\mathrm{Se_{4}}$ are governed by the magnetic state~\cite{PRL2011,PRL2015}, it is instructive to determine how the magnetic ground state evolves under pressure. For this purpose, we first performed dc and ac magnetic susceptibility measurements under high pressure. Fig. 3 (a) shows the temperature dependence of dc magnetization $\mathit{\chi}(\mathit{T})$ for Hg$\mathrm{Cr_{2}}$$\mathrm{Se_{4}}$ at various pressures up to 0.88 GPa measured with a miniature piston-cylinder cell. These data were collected upon warming up under an external magnetic field of $\mathit{H}$ = 100 Oe after zero-field cooling from room temperature.The pressure values were determined from the superconducting transition of Pb which is put inside the pressure cell together with Hg$\mathrm{Cr_{2}}$$\mathrm{Se_{4}}$. As can be seen, the FM transition temperature of Hg$\mathrm{Cr_{2}}$$\mathrm{Se_{4}}$ is gradually lowered with increasing pressure, but the FM order remains robust at 2 K up to $\sim$ 0.9 GPa. 

\begin{figure*}[t]\centering
	\centering
	\includegraphics[width=1\textwidth]{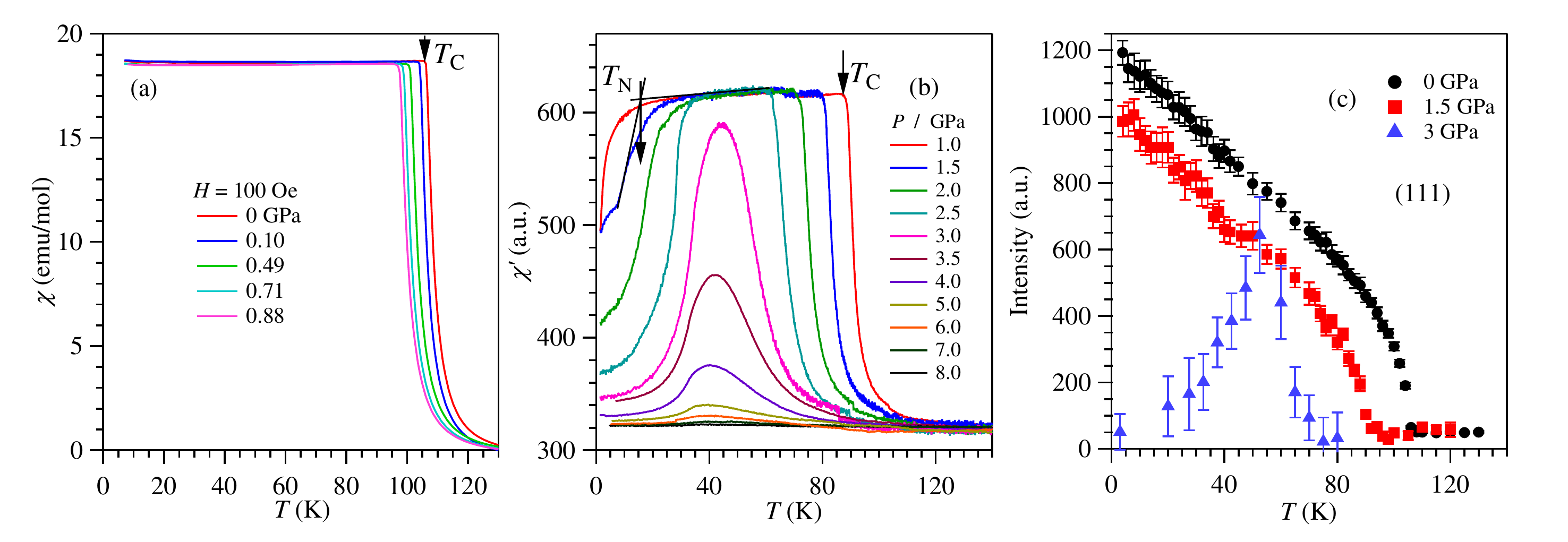}
	\caption{(color online) (a) (b) Temperature dependence of dc and ac susceptibility $\mathit{\chi}$($\mathit{T}$) of Hg$\mathrm{Cr_{2}}$$\mathrm{Se_{4}}$ under high pressures. The ferromagnetic transition temperatures $\mathit{T}_\mathrm{C}$ and the antiferromagnetic transition temperatures $\mathit{T}_\mathrm{N}$ are marked by arrows. (c) Temperature dependence of the integrated intensities of the (111) Bragg peak observed by neutron diffraction at 0, 1.5, and 3.0 GPa. The intensities between different pressures are normalized using the (111) nuclear Bragg peak intensities above $\mathit{T}_\mathrm{C}$.}
	\label{Fig. 3}
\end{figure*}

Fig. 3(b) displays the ac magnetic susceptibility $\mathit{\chi}^{\prime}(\mathit{T})$ measured from 1 to 8 GPa in a cubic anvil cell apparatus to further track the evolution of magnetic ground state under higher pressures~\cite{RSI2014}. We can see the FM $\mathit{T}_\mathrm{C}$ was continuously reduced from 106 K at ambient pressure to $\sim$ 70 K at 2.5 GPa , consistent with the resistivity data shown in Fig. 1(d). The nearly parallel shift down of $\mathit{T}_\mathrm{C}$ illustrates that the FM exchange interactions are weakened by pressure, and consistent with the resistivity data shown in Fig. 1. On the other hand, a clear drop develops below $\sim$ 10 K in $\mathit{\chi}^{\prime}(\mathit{T})$ under 1 GPa, and it becomes more pronounced and moves to higher temperatures with pressure. The sharp drop of $\mathit{\chi}^{\prime}(\mathit{T})$ appears at $\sim$ 35 K under 2.5 GPa. Similar behavior has also been observed in MnP~\cite{PRL2015Cheng}, in which the transition from FM to a double helical state induces a sharp drop of ac magnetic susceptibility as observed here. The drop of magnetic susceptibility thus suggests that the FM ground state might transform to an AF or helimagnetic state under pressure. Thus we define $\mathit{T}_\mathrm{N}$ as shown in Fig. 3(b). Since the observed drop in $\mathit{\chi}^{\prime}(\mathit{T})$ takes place at nearly the same temperature as $\mathit{T}_\mathrm{min}$ in resistivity, Fig. 1, the low-temperature upturn in resistivity should be attributed to the development of AF state. Above 2.5 GPa, the plateau in $\mathit{\chi}^{\prime}(\mathit{T})$ disappears completely and there left only a single peak anomaly, which stays nearly at the same temperature $\sim$ 40 K but its magnitude decreases gradually with pressure. The observation of broad maximum in $\mathit{\chi}^{\prime}(\mathit{T})$ of Hg$\mathrm{Cr_{2}}$$\mathrm{Se_{4}}$ at $\mathit{P}$ $\geq$ 3 GPa  is very similar with that observed in Hg$\mathrm{Cr_{2}}$$\mathrm{S_{4}}$ at ambient pressure, in which the quick drop of $\mathit{\chi}^{\prime}(\mathit{T})$ corresponds to the development of noncollinear AF or spiral magnetic order at $\mathit{T}_\mathrm{N}$ $=$ 22 K~\cite{PRB2006}. The presence of spiral rather than FM order in Hg$\mathrm{Cr_{2}}$$\mathrm{S_{4}}$ has been attributed to the enhanced further-neighbor AF  interactions relative to the nearest-neighbor FM interaction~\cite{PRB2006}. These observations thus indicate that the FM ground state of Hg$\mathrm{Cr_{2}}$$\mathrm{Se_{4}}$ is most likely replaced gradually by a noncollinear AF or spiral state similar as Hg$\mathrm{Cr_{2}}$$\mathrm{S_{4}}$. 

To verify the pressure-induced spiral magnetic ground state, we perform single-crystal neutron diffraction measurements under high pressures at HB-1 station installed at the HFIR of ORNL. Fig. 3 (c) shows the (111) Bragg peak intensities at 0, 1.5, and 3 GPa as a function of temperature. The weak signal at high temperatures above $\mathit{T}_\mathrm{C}$ represents the nuclear contribution, and the magnetic contribution develops below $\mathit{T}_\mathrm{C}$. With increasing pressure, the FM ordering $\mathit{T}_\mathrm{C}$ is decreasing which is consistent with the magnetic susceptibility measurements. Although the ac magnetic susceptibility at 1.5 GPa shows a small decrease below $\sim$ 20 K, there is no anomaly observed with the neutron diffraction measurements. This is probably because the transition temperature under pressure is slightly sample dependent or pressure mismatch for different pressure cells. With increasing pressures, the decrease of the (111) intensity corresponds to the reduction of the ordered FM moment. As can be seen, the FM ordering $\mathit{T}_\mathrm{C}$ and the magnetic moment further decrease at 3.0 GPa. Moreover, the FM component, which developed below $\sim$ 70 K, decreases sharply below $\sim$ 50 K. This behavior is consistent with our magnetic susceptibility measurements shown in Fig. 3(b). We also searched for additional magnetic peaks around (111) in the range of (HHL) with 0.8 $\leq$ H $\leq$ 1.2 and 0.8 $\leq$ L $\leq$ 1.2 below 50 K. However, no noticeable magnetic signal was observed. Although our present high-pressure neutron diffraction experiments did not provide direct evidence for the spiral magnetic order, the sharp drop of (111) intensity at 3 GPa unambiguously demonstrated the destruction of FM order by pressure, and are consistent with either disordered state or an incommensurate spiral-type AF state with short periodicity.

The stabilization of spiral magnetic order in Hg$\mathrm{Cr_{2}}$$\mathrm{Se_{4}}$ under high pressure is further supported by the comparison with Hg$\mathrm{Cr_{2}}$$\mathrm{S_{4}}$. As shown in Fig. 1, our $\mathit{\rho}(\mathit{T})$ data at 0 T and 8 T at 1 $<$ $\mathit{P}$ $<$ 3.5 GPa share some similarities with metamagnetic Hg$\mathrm{Cr_{2}}$$\mathrm{S_{4}}$, which forms a spiral magnetic order below $\mathit{T}_\mathrm{N}$ $\sim$ 22 K and can be easily converted into a FM state above $\sim$ 1 T~\cite{PRB2006,PRL2006}. For Hg$\mathrm{Cr_{2}}$$\mathrm{S_{4}}$, its resistivity first changes from a semiconductor or an insulator to metallic-like state at $\sim$ 80 K due to the presence of strong FM correlations, and then restores to a reentrant insulating behavior below 25 K when spiral magnetic order sets in. Then, the spiral order induced resistivity upturn at low temperature changes back to the metallic behavior under moderate external magnetic field when the sample is converted into FM state~\cite{PRL2006}. Thus, this comparison not only leads a strong support for the spiral order in Hg$\mathrm{Cr_{2}}$$\mathrm{Se_{4}}$ under pressure, but also highlights a very similar role of physical pressure as the chemical pressure realized via replacing Se with smaller S~\cite{PRB2008Ueda,PRB2016}.

Based on the obtained $\mathit{T}_\mathrm{C}^{\rho}$, $\mathit{T}_\mathrm{min}^{\rho}$ from resistivity and $\mathit{T}_\mathrm{C}^{\mathit{\chi}^{\prime}}$, $\mathit{T}_\mathrm{N}^{\mathit{\chi}^{\prime}}$ from magnetic susceptibility, we construct the temperature-pressure phase diagram of Hg$\mathrm{Cr_{2}}$$\mathrm{Se_{4}}$ as shown in Fig.4. The evolution of the magnetic phase transitions can be visualized more vividly in a contour plot of the ac magnetic susceptibility $\mathit{\chi}^{\prime}(\mathit{T})$ superimposed in the phase diagram. It shows explicitly the gradual suppression of the FM order followed by the emergence and subsequent growth of new AF, most likely a spiral-type magnetic order at low temperatures. From the phase diagram, we can also see that the FM order and the spiral magnetic order are merged at $\sim$ 4 GPa, and the phase boundary between paramagnetic (PM) semiconducting and spiral insulating regions cannot be distinguished at higher pressures. Moreover, the phase diagram also highlights the fact that at ambient pressure Hg$\mathrm{Cr_{2}}$$\mathrm{Se_{4}}$ should be located close to a critical point where the spiral magnetic state strongly compete with the FM ground state.

Our magnetic susceptibility and neutron diffraction measurements under high pressure provide direct evidences that can link the resistivity upturn below $\mathit{T}_\mathrm{min}$ or the development of insulating behavior in $\mathit{n}$-type Hg$\mathrm{Cr_{2}}$$\mathrm{Se_{4}}$ with pressure destabilization of FM ground state. According to a recent theoretical study on the phase diagram of chromium spinels~\cite{PRX2017}, Hg$\mathrm{Cr_{2}}$$\mathrm{Se_{4}}$ is located inside the FM region at ambient pressure due to the presence of dominant nearest-neighbor (NN) FM interaction $\mathit{J_\mathrm{1}}$, negligibly second NN FM $\mathit{J_\mathrm{2}}$ $\sim$ 0.0014$\mathit{J_\mathrm{1}}$, and sizable third NN AF $\mathit{J_\mathrm{3}}$ $\sim$ -0.109$\mathit{J_\mathrm{1}}$. In the phase diagram of $\mathit{J_\mathrm{3}}$/$|$$\mathit{J_\mathrm{1}}$$|$ versus $\mathit{J_\mathrm{2}}$/$|$$\mathit{J_\mathrm{1}}$$|$, Hg$\mathrm{Cr_{2}}$$\mathrm{Se_{4}}$ is close to the boundary between FM and spiral state and the enhancement of either $\mathit{J_\mathrm{2}}$ or $\mathit{J_\mathrm{3}}$ relative to $\mathit{J_\mathrm{1}}$ can drive it closer or into the spiral magnetic state. In the present case, the reduction of Cr-Cr distances under pressure that can effectively enhance both $\mathit{J_\mathrm{2}}$ and $\mathit{J_\mathrm{3}}$ should thus favor the spiral magnetic order. As a matter of fact, analyses of low-temperature specific heat have suggested the presence of AF contributions inside the FM ground state at ambient pressure~\cite{JLTP2013}. Apparently, the application of high pressure effectively tips the balance of competition between NN FM and thrid NN AF interactions and should eventually stabilizes a spiral magnetic ground state. Because the chemical potential of half-metallic Hg$\mathrm{Cr_{2}}$$\mathrm{Se_{4}}$ is very small, a deviation from perfect ferromagnetism or even the stabilization of spiral magnetic order  under pressure would reduce the $\mathit{s}$-$\mathit{d}$ exchange splitting and destroy the metallic state by opening a gap between the Hg-6$\mathit{s}$ and Se-4$\mathit{p}$ bands~\cite{PRL2011,PRL2015}.

Our high pressure measurements also highlight that $\mathit{n}$-type Hg$\mathrm{Cr_{2}}$$\mathrm{Se_{4}}$ is a very unique single-valent FM metallic system which involves the strong coupling between the lattice, electronic, and magnetic degrees of freedom. It is so sensitive that the external pressure and magnetic field can easily tune the ground state between the spiral insulating state and FM metallic state. Through switching between different magnetic ground states, an extremely large MR as high as $\mathrm{10^{11}}$ $\%$ can  be achieved due to the strong coupling between the charge and spin degrees of freedom. By utilizing the chemical substitutions, if we can push the FM transition to room temperature, this feature can be used to harvest applicable spin electronic devices, such as spin valve or the memory storage devices.

\begin{figure}[t]
	\centering
	\includegraphics[width=0.45\textwidth]{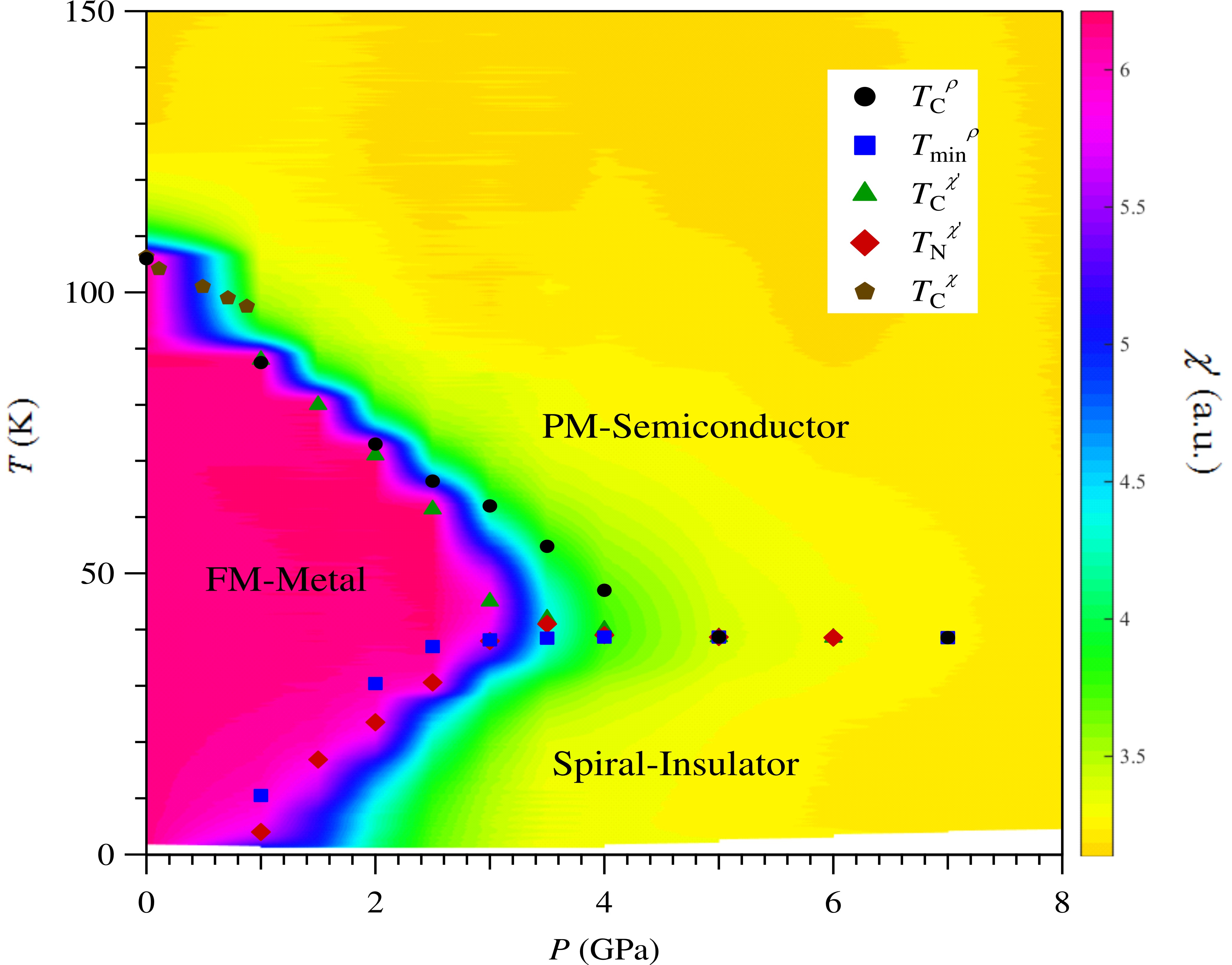}
	\caption{(color online) Temperature-pressure phase diagram of $\mathit{n}$-type Hg$\mathrm{Cr_{2}}$$\mathrm{Se_{4}}$. The ferromagnetic order temperatures ($\mathit{T}_{\mathrm C}$, black filled circle, green triangle and brown pentagon), the upturn in resistivity ($\mathit{T}_\mathrm{min}$, blue square), and the spiral magnetic order temperatures ($\mathit{T}_{\mathrm N}$, red rhombuse) as a function of hydrostatic pressure. }
	\label{Fig. 4}
\end{figure}

In summary, we have elucidated that pressure and magnetic field can tip the balance between the FM and AF interactions in $\mathit{n}$-type Hg$\mathrm{Cr_{2}}$$\mathrm{Se_{4}}$ in an opposite way. The application of high pressure suppresses the FM order by enhancing the AF exchange coupling, and gradually stablizes the spiral magnetic and insulating ground state. On the other hand, magnetic field can easily modify the magnetic interactions to switch the ground state back to FM metallic state. As such, the magnetic competition induces prominent colossal magnetoresistance $\sim$ $\mathrm{10^{11}}$ $\%$ at low temperatures. Thus, our present work provides a means for realizing novel state where the extremely large magnetoresisitance can be obtained via switching between two distinct electronic ground states in a single-valent system.

\emph{Acknowledgments.}---This work is supported by the National Key R$\&$D Program of China (Grant No. 2018YFA0305700 and Grant No. 2017YFA0302901), the National Natural Science Foundation of China (Grant No. 11888101, 11574377, 11834016, 11874400, 61425015, 11774399), the Strategic Priority Research Program and Key Research Program of Frontier Sciences of the Chinese Academy of Sciences (Grant Nos. XDB25000000, XDB07020100 and Grant No. QYZDB-SSW-SLH013), and Beijing Natural Science Foundation (Z180008). A portion of this research used resources at the High Flux Isotope Reactor, a DOE Office of Science User Facility operated by the Oak Ridge National Laboratory. Y.Y.J. and J.P.S. acknowledge support from the China Postdoctoral Science Foundation and the Postdoctoral Innovative Talent program.

\bibliography{refs}

\end{document}